\documentclass[reqno]{llncs}
\usepackage{framed}
\usepackage{url}
\usepackage[ruled,vlined,linesnumbered]{algorithm2e}
\usepackage{comment}
\usepackage{tikz}
\usetikzlibrary{positioning,quotes}
\usepackage[utf8]{inputenc}
\usepackage{array}
\usepackage{color, colortbl}
\usepackage[T1]{fontenc}
\usepackage{todonotes}
\usepackage{float}
\usepackage{comment}
\usepackage[square,numbers]{natbib}
\usepackage{lineno}

\makeatletter
\renewcommand\@biblabel[1]{#1.}
\makeatother

\usepackage{soul}
\usepackage{multirow}
\usepackage[export]{adjustbox}
\usepackage{amsfonts} 

\usepackage[leqno]{amsmath}
\usepackage{threeparttable}
\usepackage{amsmath}
\usepackage{multirow}
\usepackage{threeparttable}
\usetikzlibrary{arrows,shapes}
\usepackage{algorithmicx}
\usepackage{cleveref}

\newcommand{\DFSGDEXT}{\textbf{\scriptsize{DFS-GREEDY-PRUNNING }}}
\newcommand{\BFSGDEXT}{\textbf{\scriptsize{BFS-GREEDY-PRUNNING }}}
\newcommand{\BB}{\textbf{\scriptsize{BACKBONE-BASED }}}
\newcommand{\AM}{\textbf{\scriptsize{AVERAGE-DEGREE }}}
\newcommand{\SEEDF}{\textbf{\scriptsize{GET-POTENTIAL-SEEDS }}}
\newcommand{\DIFFU}{\textbf{\scriptsize{DIFFUSION }}}
\newcommand{\MOSTINV}{\textbf{\scriptsize{MOST-INACTIVE-VERTEX }}}
\newcommand{\MOSTINVS}{\textbf{\scriptsize{GET-MOST-INACTIVE-VERTICES }}}
\newcommand{\EXTR}{\textbf{\scriptsize{PRUNING }}}
\newcommand{\HPT}{\textbf{\scriptsize{GET-POTENTIAL-NODE-BY-BFS-TREE }}}
\newcommand{\G}{$G(V,E)$}
\newcommand{\scap}{$\hat{S}$}
\newcommand{\PNAME}{MINFS}

\tikzset{
    process1/.style={rectangle, minimum width=8cm, minimum height=1cm, text width=10cm,text centered, draw=black},
    process2/.style={rectangle, minimum width=4cm, minimum height=1cm, text width=5cm,text centered, draw=black},
    decision/.style={diamond, text centered, draw=black, aspect=2, inner xsep=-2mm},
    stop/.style={rectangle, rounded corners, minimum width=2cm, minimum height=1cm,text centered, draw=black},
    decision1/.style={diamond, text centered, draw=black, aspect=2, inner xsep=-2mm},
    stop/.style={rectangle, rounded corners, minimum width=2cm, minimum height=1cm,text centered, draw=black},
    arr/.style={thick,-stealth}
    }
    
\newcounter{cases}
\newcounter{subcases}[cases]

\usepackage{geometry}
\usepackage{multirow,tabularx}

\newcolumntype{Y}{>{\centering\arraybackslash}X}

\DeclareMathOperator*{\argmax}{argmax}

\urldef{\mails}\path|19mcpc06@uohyd.ac.in,askcs@uohyd.ac.in,sdbcs@uohyd.ac.in|

\SetKwComment{Comment}{$\triangleright$\ }{}

\title{Heuristics for Influence Maximization with Tiered Influence and Activation thresholds}

\author{Rahul Kumar Gautam$^{0009-0009-7693-3863}$ \and Anjeneya Swami Kare$^{0000-0003-3644-4802}$ \and  Durga Bhavani S.$^{0000-0003-4413-0328}$}
\authorrunning{F. Author et al.}
\institute{School of Computer and Information Sciences, \\University of Hyderabad,\\ Hyderabad, India\\ \mails}

\tikzstyle{vertex}=[circle,fill=black!25,minimum size=12pt,inner sep=0pt]
\tikzstyle{selected vertex} = [vertex, fill=red!60]
\tikzstyle{edge} = [draw,thick,-]
\tikzstyle{weight} = [font=\small]
\tikzstyle{selected edge} = [draw,line width=3pt,-,red!50]
\tikzstyle{ignored edge} = [draw,line width=3pt,-,black!20]

\begin{document}
\raggedbottom
\maketitle
\begin{abstract}
The information flows among the people while they communicate through social media websites. Due to the dependency on digital media, a person shares important information or regular updates with friends and family. The set of persons on social media forms a social network. Influence Maximization (IM) is a known problem in social networks. In social networks, information flows from one person to another using an underlying diffusion model. There are two fundamental diffusion models: the Independent Cascade Model (ICM) and the Linear Threshold Model (LTM). In this paper, we study a variant of the IM problem called Minimum Influential Seeds (\PNAME{})  problem proposed by~\citeauthor{2023tieredInf}\cite{2023tieredInf}. It generalizes the classical IM problem with LTM as the diffusion model. Compared to IM, this variant has additional parameters:  the influence threshold for each node and the propagation range. The propagation range is a positive integer that specifies how far the information can propagate from a node. A node on the network is not immediately influenced until it receives the same information from enough number of neighbors (influence threshold). Similarly, any node does not forward information until it receives the same information from a sufficient number of neighbors (activation threshold). Once a node becomes activated, it tries to activate or influence its neighbors. The \PNAME{} problem aims to select the minimum number of initial spreader nodes such that all nodes of the graph are influenced.

In this paper, we extend the study of the \PNAME{} problem. We propose heuristics that construct seed sets based on the average degree of non-activated nodes, closest first, and backbone-based heaviest path. We have also proposed a pruning technique that further reduces the size of the seed sets. We have implemented the existing heuristics and the proposed heuristics. We have done extensive experimentation on $18$ real-world data sets. The proposed heuristics give improved seed sets compared to the existing heuristics.

\end{abstract}
\section{Introduction}\label{sec:intro}

In this digital world, people get news or information digitally on their gadgets. Due to the advantages of social media, human beings are rapidly adopting social media in their daily life. However, there are some pros and cons of social media. Some disadvantages are rumor-spreading, privacy-related issues, data theft, etc. 
Nevertheless, getting important information about what is happening in society becomes essential. As we have seen during the COVID-19 pandemic, the government has to make people aware of the pandemic and related safety measures. Almost all countries' governments use social media to run awareness campaigns as it saves the time and effort of the government. 

On account of the enormous applications of social media, the discussion on how information propagates on social media networks becomes very important. The influence maximization problem is related to information propagation and maximizing the influenced people in social networks. A node in the social network is said to be influenced by a message when it starts believing the message. On the other hand, a node is said to be activated when it starts forwarding (spreading) the message to its neighbors. 

~\citeauthor{kempe2003maximizing}\cite{kempe2003maximizing} introduced the Influence Maximization (IM) problem. The IM problem is also called the Target Set Selection (TSS) problem. Using the Linear Threshold Model (LTM) of diffusion, there are primarily two variants of the TSS problem: the maximization version and the minimization version. For the maximization version, input is a graph $G=(V, E)$ and a positive integer $k$, and the problem asks to compute a target set (seed set) $S \subseteq V$ of size at most $k$ that activates the maximum number of vertices. For the minimization version, input is a graph $G$ and an integer input $\ell$, and the problem asks to compute a target set (seed set) $S \subseteq V$ of the minimum size that activates at least $\ell$ vertices. If $\ell = |V|$, then the problem asks to compute a target set $S \subseteq V$ of the minimum size that activates all the vertices of the graph. 

~\citeauthor{cordasco2018evangelism}\cite{cordasco2018evangelism} studied a variant of the TSS problem, which they called the Perfect Evangelizing Set (PES) problem. For the PES problem, input is a graph $G$, influence and activation (evangelizing ) thresholds $t_I, t_A: V \rightarrow \{0, 1, 2, \dots, \}$ and the problem asks to compute a target set (seed set) $S \subseteq V$ of minimum size that influences all the vertices of the graph. ~\citeauthor{cordasco2019active}\cite{cordasco2019active} also introduced a problem called the Perfect Awareness (PA) problem, which is a specialization of the PES problem. In the PA problem $t_I(v) = 1, \forall v\in V$.

~\citeauthor{2023tieredInf}\cite{2023tieredInf} proposed a variant of the TSS problem, which they call the Minimum Influential Seeds (\PNAME{}) problem, which is a generalization of the PES problem. Compared to the PES problem, the \PNAME{} problem has an additional input parameter called the propagation range $p\in \mathbb{Z}^+$. The propagation range indicates how far the information propagates from one node to another node. The \PNAME{} problem with $p = Diameter(G)$ is equivalent to the PES problem. For the experimentation~\citeauthor{2023tieredInf}\cite{2023tieredInf} used two input parameters $\theta$ and $\alpha$ such that $0 < \theta \leq \alpha \leq 1$. For each vertex $u \in V$, they set the influence threshold $t_I(u) = \theta.deg(u)$ and the activation threshold $t_A(u) = \alpha.deg(u)$. In the \PNAME{} problem, the information can flow from the sources (initial spreaders) up to $p \in \mathbb{Z^+}$ distance. If a vertex $u$ receives information from at least  $t_I(u)$ neighbors, $u$ becomes influenced. Likewise, If a vertex $u$ receives information from at least  $t_A(u)$ neighbors, $u$ becomes activated and forwards information to the neighbors. An initial spreader can not activate or contribute to activating vertices at more than $p$ distance. The set of initial spreaders is called a seed set. The objective is to find the set of initial spreaders (seed set) of minimum size, which influences all the vertices of the graph.

~\citeauthor{2023tieredInf}\cite{2023tieredInf} proposed minimization and maximization variants of the Influential Seeds problem. They have proposed heuristics for the \PNAME{} problem. In this paper, we extend the study of the \PNAME{} problem.  We propose three heuristics and a pruning strategy to improve the solutions obtained by the heuristics. We have the following results:      

\begin{enumerate}
    \item[-] The first heuristic picks the average number of highly influential inactive vertices for the seed set in each iteration. It improves the quality of the result and running time compared to existing heuristics for the problem.
    \item[-] The second heuristic finds the closest highly influential seed vertex from the seed set. 
    \item[-] The third heuristic, a backbone-based heuristic, finds the dominant path and selects vertices from the dominant path for the seed set.
    \item[-] The proposed pruning technique is applied to solutions returned by the existing and as well as the proposed heuristics. The pruning technique improves the quality of the solutions.
\end{enumerate}


The whole paper is organized as follows. In section~\ref{sec:relwork}, we present related and recent studies on the \PNAME{} problem, such as influence maximization, perfect awareness problem, and the target set selection problem. Section~\ref{sec:proposedAlgo} covers our proposed heuristics. In section~\ref{sec:resultanddiscussion}, we analyze the performance of algorithms and present results on the real datasets. The final section of the article concludes the paper.

\section{Related Work}\label{sec:relwork}
Influence Maximization (IM) in social networks is an essential area of research due to its applications in business advertisements, viral marketing, and campaigning. The IM problem is also known as the Target Set Selection (TSS) problem.~\citeauthor{kempe2003maximizing}\cite{kempe2003maximizing} introduced the IM problem. They proved that the IM problem is NP-hard and proposed experimental algorithms for the problem. The greedy algorithm by~\citeauthor{kempe2003maximizing}\cite{kempe2003maximizing} guarantees approximation $1-(1/e)$ ($e$ is logarithmic base). For the TSS problem based on the decreasing cascade model, the $1-(1/e)-\epsilon$-approximation algorithm was studied by~\citeauthor{kempe2003maximizing}\cite{kempe2003maximizing}.~\citeauthor{chen2009approximability}\cite{chen2009approximability} studied the hardness of the TSS problem and proved that the problem is hard to approximate within a poly-logarithmic factor. There are two fundamental diffusion models: the Independent Cascade Model (ICM) and the Linear Threshold Model (LTM). A vertex influences its neighbors with some probability in an independent cascade model. In the linear threshold model, each vertex is activated or influenced if the vertex has a number of active neighbor spreaders greater than or equal to the threshold value of the vertex.~\citeauthor{cordasco2018evangelism}\cite{cordasco2018evangelism} studies evangelism in social networks based on the linear threshold model, in which each vertex has an influence threshold and activation threshold. When a vertex receives information from an influence threshold number of neighbors, it becomes influenced. For a vertex to become activated, the vertex should have at least an activation threshold number of activated neighbors. Later,~\citeauthor{cordasco2019active}\cite{cordasco2019active} presented a Perfect Awareness (PA) problem on the linear threshold model in which the influential threshold for each vertex is considered one. The heuristics ~\cite{2023RahulCent,pereira2021effective} of the PA problem are proposed.

The $k$-center problem~\cite{k-center}, PA problem~\citep{cordasco2019active}, evangelism in social networks~\citep{cordasco2018evangelism}, graph burning problem~\citep{Bonato2016burn}, opinion maximization~\citep{alla2023opinion}, target  influence maximization in competitive ~social networks\cite{2023targetedCompt}  under the independent cascade model, and rumor minimization~\citep{rumorMinimization} are related problems to the \PNAME{} problem.
The experimental works on the problems PA, graph burning number, and opinion maximization are proposed in ~\citep{2023RahulCent,pereira2021effective,kazemzadeh2022influence,gautam2022faster,geneticGraphBurning,alla2023opinion,Simon2019heuristic}.

In the real-life scenario, the information does not flow continuously in the social media networks. Over time, the propagation of information or advertisements in social media networks gets exhausted due to people's waning interest. So, the distance traversed by the information in social media networks needs to be addressed.~\citeauthor{2023tieredInf}\cite{2023tieredInf} study the issue in the diffusion process and introduce a significant constraint as propagation range (information can traverse distance up to the propagation range from the initial spreaders). Due to the importance of propagation range in real scenarios, we study the \PNAME{} problem and propose three heuristics for the \PNAME{} problem.

\section{ Problem  Definition}
In a given graph \G{} and a set $S \subseteq V$, initially, only the vertices of the set $S$ are influenced and activated. 
The variable $p\in \mathbb{Z}^+ $ denotes the propagation range. Let $A$ be the set of activated vertices initially $A=S$.  Initially, each activated vertex $u \in S $ can send the information up to $p$ distance. In the diffusion process,  If $ |N_A(v)| \ge \alpha * deg(v)$ where  $N_A(v)$ is a set of active neighbors of $v$,  then $v$ is added to $A$ with the condition that $v$ is activated by a smallest subset  $S'\neq \phi$. 
Similarly, if $ |N_A(v)| \ge \theta*deg(v) $, then $v$ becomes influenced. We repeat the above steps until all the vertices of the set $V$ are influenced. The objective is to find the minimum set of seed nodes ( seed set $S$ of minimum size).

\section{Proposed Algorithms}
In this section, we propose three heuristics for the \PNAME{} problem. The approach constitutes two significant steps. In the first step, the heuristics compute a potential seed set iteratively.  In the second step, the potential seed set is pruned to obtain the smallest seed set.

\label{sec:proposedAlgo}
The tricky part of solving the \PNAME{} problem is to find the smallest seed set $ S \subseteq V$. In this paper, the heuristic algorithms construct the seed set $S$ with criteria such that the algorithm influences the set of vertices $V$ through the diffusion process. The diffusion process is implemented using the Breath First Search approach~\cite{2023tieredInf}.

\subsection{Average Degree Heuristic}
\label{subsec:algofirst}
The high-level idea of the proposed \SEEDF method as given in Algorithm~\ref{algo:Mainfunctionoflistofselctedvertices} is as follows. The \SEEDF takes input parameters of graph $G$, propagation range $p$, influence threshold $\theta$, and activation threshold $\alpha$. Initially, all the vertices of \G{} are inactive, so initialize activated set $A=\phi$, influenced set $I= \phi$, and initial spreader list \scap $ = [ \; ]$. \scap{} contains the potential seed nodes and is treated as a list since the order in which the nodes are added to \scap{} is important for \EXTR step.
Repeat the following steps ($1$ and $2$) if a non-influenced node exists in the graph $G$. 
\begin{enumerate}
    \item Select a potential list of spreaders $L$  by calling the \MOSTINVS method as given in Algorithm~\ref{algo:mostinactivevertices}.
    \item For all vertices of $u \in L$.
    \begin{enumerate}
        \item Append $u$ to \scap.
        \item Add $u$ to the activation set $A$.
        \item \DIFFU function updates activation set $A$ and influence set $I$.
    \end{enumerate}
    
\end{enumerate}

As given in Algorithm~\ref{algo:mostinactivevertices}, \MOSTINVS method returns $\lceil \frac{n''}{n'} \rceil$ number of highly influential vertices where $n'$ is the number of inactive vertices in the graph $G$, and $n''$ is the sum of the inactive-degree of inactive vertices ( inactive-degree $d_{V\backslash A}(v)$ means the number of inactive neighbors of $v$). The number $\lceil \frac{n''}{n'} \rceil$ indicates the average degree of the graph  $G(V\backslash A,E)$. In each iteration  (in Algorithm~\ref{algo:mostinactivevertices}, lines 6-10), select a vertex $w\notin L$ with maximum $d_{V\backslash A}(w)$ value and append $w$  to $L$. 

\begin{figure}
\centering
\hspace{-9cm}
\begin{tikzpicture}[
node distance =10mm and 5mm,
C/.style = {circle, draw, minimum size=1em},
S/.style = {circle,fill=red!25,draw, minimum size=1em},
every edge quotes/.style = {auto, font=\footnotesize, sloped}
                    ] 

\node (1) [S, ] {1};
\node (0) [C, above= of 1] {0};
\node (2) [C,  above right=of 1] {2};
\node (3) [C,  below right=of 2] {3};
\node (4) [S, above right=of 2] {4};
\node (5) [S, dashed,  right=of 2] {5};
\node (6) [S,  below right=of 3] {6};
\node (7) [C,  below right=of 1] {7};
\node (8) [S,  above right=of 5] {8};
\node (9) [S, right=of 5] {9};
\node (10) [S,  right=of 3] {10};
\node (11) [C,  right=of 9] {11};
\node (12) [C,  right=of 10] {12};
\node (13) [C, right=of 8] {13};
%
\foreach \x in {0}
    \foreach \y in {2,4}
    {
      \draw (\x)--(\y) ;
      }
      
\foreach \x in {1}
    \foreach \y in {2,5,3,7}
    {
      \draw (\x)--(\y) ;
      }
\foreach \x in {2}
    \foreach \y in {3,5,7,4}
    {
      \draw (\x)--(\y) ;
      }

\foreach \x in {3}
    \foreach \y in {7,9,10,6}
    {
      \draw (\x)--(\y) ;
      }

\foreach \x in {5}
    \foreach \y in {10,9,4,8}
    {
      \draw (\x)--(\y) ;
      }

\foreach \x in {9}
    \foreach \y in {13,11,12}
    {
      \draw (\x)--(\y) ;
      }
    \draw(5.5,3)--(5.5,-1);
\end{tikzpicture}

\vspace{-5cm}
\hspace{3cm}
\begin{tikzpicture}[
node distance =10mm and 5mm,
C/.style = {circle, draw, minimum size=1em},
S/.style = {circle,fill=red!25,draw, minimum size=1em},
every edge quotes/.style = {auto, font=\footnotesize, sloped}
                    ] 

\node (1) [S, ] {1};
\node (0) [C, above= of 1] {0};
\node (2) [C,  above right=of 1] {2};
\node (3) [C,  below right=of 2] {3};
\node (4) [S, above right=of 2] {4};
\node (5) [C, dashed,fill=red!60, right=of 2] {5};
\node (6) [S,  below right=of 3] {6};
\node (7) [C,  below right=of 1] {7};
\node (8) [S,  above right=of 5] {8};
\node (9) [S, right=of 5] {9};
\node (10) [S,  right=of 3] {10};
\node (11) [C,  right=of 9] {11};
\node (12) [C,  right=of 10] {12};
\node (13) [C, right=of 8] {13};
%
\foreach \x in {0}
    \foreach \y in {2,4}
    {
      \draw (\x)--(\y) ;
      }
      
\foreach \x in {1}
    \foreach \y in {2,5,3,7}
    {
      \draw (\x)--(\y) ;
      }
\foreach \x in {2}
    \foreach \y in {3,5,7,4}
    {
      \draw (\x)--(\y) ;
      }

\foreach \x in {3}
    \foreach \y in {7,9,10,6}
    {
      \draw (\x)--(\y) ;
      }

\foreach \x in {5}
    \foreach \y in {10,9,4,8}
    {
      \draw (\x)--(\y) ;
      }

\foreach \x in {9}
    \foreach \y in {13,11,12}
    {
      \draw (\x)--(\y) ;
      }
\end{tikzpicture}
\caption{The spreader list of nodes(in light red color) is $\hat{S} = [5,1,4,9,8,10,6]$ and influenced vertices are in white (shown in the left). The \EXTR removes extra spreader node 5(in red) and returns $[1,4,9,8,10,6]$ because $5$ is activated by $4,8,9, 10$ (shown on the right). }
\label{fig:pruning}
\end{figure}
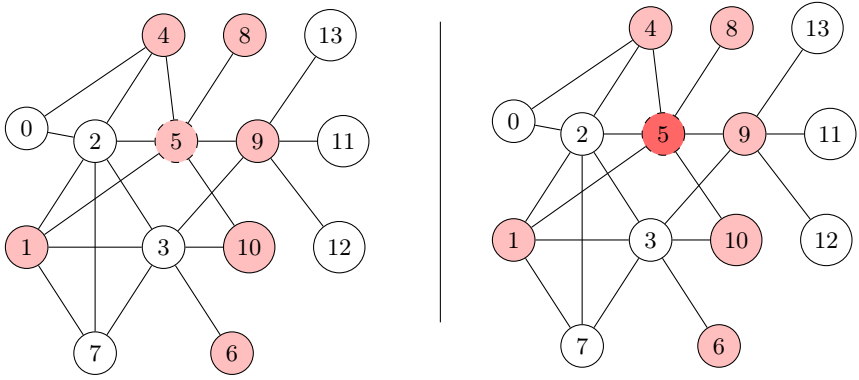

We use \EXTR to refine the spreader list \scap{} obtained by the above steps ($1$ and $2$).
The \EXTR method, as given in algorithm~\ref{algo:extra-removal}, removes the extra spreaders from the list \scap{} and returns the final seed set $S$. The importance of pruning is illustrated in Figure~\ref{fig:pruning}.

\begin{algorithm}[ht]
	\DontPrintSemicolon
	\SetKwInOut{Input}{Input}\SetKwInOut{Output}{Output}
	\Input   {Graph \G, propagation range $p$, influential and activation thresholds ( $\theta$ and $\alpha$).}
	\Output  {Potential spreader list \scap}
	\SetKwFunction{FMain}{\SEEDF}
	\SetKwProg{Fn}{Function}{}
	\Fn{\FMain{$G,p,\theta, \alpha$}}{
		\Begin{
            $A \gets \phi $\;
            $I \gets \phi $\;
             \scap $ \gets [\;]$\;
            \While{$|I|\neq |V(G)|$}{
                  $L \gets  \textbf{{\MOSTINVS(G,A)}}$\;
                  \For{ $u \in L $}{
                     $ append(\hat{S},u)$\;
                     $ A = A \cup \{ u \}$\;
                     $[I,A] \gets \textbf{\DIFFU}(G,u, p, A, \alpha,\theta) $\;
                     \If{$|I|=|V(G)|$}
                     {
                      break\;
                     }
                  }
            }
             \KwRet{\scap}
		}		
	}
	\caption{Get potential seed nodes.}
	\label{algo:Mainfunctionoflistofselctedvertices}
\end{algorithm}

\begin{algorithm}[ht]
	\DontPrintSemicolon
	\SetKwInOut{Input}{Input}\SetKwInOut{Output}{Output}
	\Input   {Graph \G{} and activated set $A$.}
	\Output  {List of vertices $L$ which have highly inactive neighbors.}
	\SetKwFunction{FMain}{\MOSTINVS}
	\SetKwProg{Fn}{Function}{}
	\Fn{\FMain{$G, A$}}{
		\Begin{
            $L \gets [ \text{ } ]$\;
            $n' = |\{v: v\in V\backslash A\}|$ \;
            $n'' = \underset{u \in V \backslash A} \sum d_{V\backslash A}(u)$ \;
            \tcp{$d_{V\backslash A}(u)$ number of inactive neighbors of $u$}
            \For{$1 \text{ to } \lceil \frac{n''}{n'} \rceil $}{
                $w = \underset{v}{ \argmax} \{d_{V\backslash A}(v) : v \in V\backslash L \}$\;
             \If{$w \neq -1 $}{
             $append(L,w)$\;
             }
            }
 
             \KwRet{$L$}
		}		
	}
	\caption{List of most inactive vertices.}
	\label{algo:mostinactivevertices}
\end{algorithm}
The \EXTR method reverses the list \scap. For each vertex $u \in \hat{S}$, check whether the diffusion process can influence all vertices of the graph by spreader list $\hat{S} \backslash \{u\}$. If yes, remove the vertex $u$ from \scap; otherwise, $u$ must be present in the spreader list \scap. The \EXTR method removes the extra spreader nodes from \scap{} and returns the final seed set $S$. The \EXTR method gives results based on the order of vertices selected as the initial spreader. So, if we apply \EXTR on two lists of equal sizes but with different sequences of spreader lists having the same elements then \EXTR may return two  reduced lists of different sizes  of initial spreaders in both cases. Therefore, \EXTR performance depends on how we construct the potential spreader list \scap.

\begin{algorithm}[!htb]
	\DontPrintSemicolon
	\SetKwInOut{Input}{Input}\SetKwInOut{Output}{Output}
	\Input  {Graph \G, propagation step $p$, potential spreader list \scap, influential and activation thresholds ( $\theta$ and $\alpha$).}
	\Output  {Final seed set $S$.}
	\SetKwFunction{FMain}{\EXTR}
	\SetKwProg{Fn}{Function}{}
	\Fn{\FMain{$G, p, \alpha, \theta$, \scap}}{
		\Begin{
            $S \gets \phi $\;
            $\hat{S}' \gets reverse(\hat{S})$\;
            \For{$u \in \hat{S}' $}{
               $S \gets \hat{S} \backslash \{ u \}$\;
               $[A,I] \gets \DIFFU(G,p,\alpha,\theta,S)$\;
               \If{$|I|= |V(G)|$}{
                $\hat{S} \gets S$\;
               }
            }
             \KwRet{$S$}
		}		
	}
	\caption{Final seed set $S$ after pruning the list \scap.}
	\label{algo:extra-removal}
\end{algorithm}
\subsection{Closest First Heuristic}\label{subsec:algoclosest}
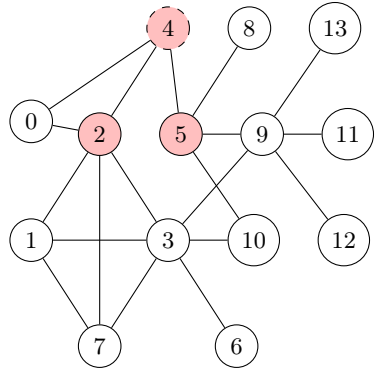
\begin{figure}[H]
\centering
\begin{tikzpicture}[
node distance =10mm and 5mm,
C/.style = {circle, draw, minimum size=1em},
S/.style = {circle,fill=red!25,draw, minimum size=1em},
every edge quotes/.style = {auto, font=\footnotesize, sloped}
                    ] 

\node (1) [C, ] {1};
\node (0) [C, above= of 1] {0};
\node (2) [S,  above right=of 1] {2};
\node (3) [C,  below right=of 2] {3};
\node (4) [S, dashed, above right=of 2] {4};
\node (5) [S, right=of 2] {5};
\node (6) [C,  below right=of 3] {6};
\node (7) [C,  below right=of 1] {7};
\node (8) [C,  above right=of 5] {8};
\node (9) [C, right=of 5] {9};
\node (10) [C,  right=of 3] {10};
\node (11) [C,  right=of 9] {11};
\node (12) [C,  right=of 10] {12};
\node (13) [C, right=of 8] {13};
%
\foreach \x in {0}
    \foreach \y in {2,4}
    {
      \draw (\x)--(\y) ;
      }
      
\foreach \x in {1}
    \foreach \y in {2,3,7}
    {
      \draw (\x)--(\y) ;
      }
\foreach \x in {2}
    \foreach \y in {3,7,4}
    {
      \draw (\x)--(\y) ;
      }

\foreach \x in {3}
    \foreach \y in {7,9,10,6}
    {
      \draw (\x)--(\y) ;
      }

\foreach \x in {5}
    \foreach \y in {10,9,4,8}
    {
      \draw (\x)--(\y) ;
      }

\foreach \x in {9}
    \foreach \y in {13,11,12}
    {
      \draw (\x)--(\y) ;
      }
\end{tikzpicture}

\caption{The first seed vertex is $2$ and the second most inactive vertex within two hops is $5$. We see here $2$ and $5$ are activating vertex $4$.}
\label{fig:closestfirst}
\end{figure}

Suppose  $d_{V\backslash A}(u)$ ( where $d_{V\backslash A}(u)$ is the number of inactive neighbors of the vertex $u$) and $d_{V\backslash A}(v)$  are the top two highest values in  the graph $G(V\backslash A, E)$, where $d(u,v)\le 2$. Both $u$ and $v$ together can activate and influence more vertices.  As given in Fig.~\ref{fig:closestfirst}, the vertices  $2$ and $5$  activate and influence $4$ immediately. Therefore, the next seed vertex $w$ is selected within two hops from the \scap{} nodes with a maximum $d_{V\backslash A}(w)$ value. 

Initially, the spreader list \scap{} is empty. Find a  vertex $w$ with the most inactive neighbors within two hops from all vertices in \scap{} and add $w$ to \scap. If no inactive vertex exists within two hops from the nodes in \scap{} and the whole graph \G{} is not influenced, a vertex with the most inactive neighbors is added to \scap. On each addition of a seed vertex to list \scap, the diffusion process diffuses the information in graph $G$ and updates influenced and activated sets $I$ and $A$. The process stops when all vertices are influenced. After finding list \scap, the \EXTR process removes extra spreaders from the list \scap{} and returns the final seed set $S$. 

\subsection{Backbone-Based Heuristic}
\label{subsec:algotwo}
As we saw in the previous algorithm, given in the algorithm~\ref{algo:Mainfunctionoflistofselctedvertices} where \MOSTINVS method returns highly influential list $L$. We add each vertex of list $L$ one by one to the spreader list \scap. Instead of adding all vertices from $L$ to spreader list \scap, append-only a vertex $u$ from list $L$ with the most inactive surrounding vertices. The reason for selecting a vertex $u$ is that $u$ belongs to a dense sub-graph of inactive vertices.      

As given in algorithm~\ref{algo:SeedNodesBYBFSTree}, the \HPT method finds a tree simultaneously from the list of roots $L$ by assuming the weight on the vertices as the number of inactive neighbor vertices in the graph $G(V\backslash A, E)$. It returns a root $w\in L$ associated with the heaviest BFS tree. The vertex $w$ is included in \scap{} and $A$, and the \DIFFU process marks vertices as activated or influenced. The process stops when all the vertices of \G{} become influenced. The \EXTR method removes the extra spreaders from the list \scap{} and returns the seed set $S$.

\begin{algorithm}[ht]
	\DontPrintSemicolon
	\SetKwInOut{Input}{Input}\SetKwInOut{Output}{Output}
	\Input   {Graph \G, propagation step $p$, influential and activation thresholds ( $\theta$ and $\alpha$).}
	\Output  {Final spreader list $S$}
	\SetKwFunction{FMain}{BACKBONE-BASED-SEED-NODES}
	\SetKwProg{Fn}{Function}{}
	\Fn{\FMain{$G,p,\theta, \alpha$}}{
		\Begin{
            $S \gets \phi $\;
            $A \gets \phi $\;
            $I \gets \phi $\;
            $\hat{S} \gets [\;]$\;
            \While{$I\neq V(G)$}{
                  $L \gets  \textbf{{\MOSTINVS(G,A)}}$\;
                  $u \gets \HPT(G,A,L)$\;
                     $ append(\hat{S},u)$\;
                     $ A = A \cup \{ u \}$\;
                     $[I,A] \gets \textbf{\DIFFU}(G,u, p, A, \alpha,\theta) $\;
            }
             \KwRet{$S$}
		}		
	}
	\caption{Get potential seeds by using \HPT method.}
	\label{algo:SeedNodesBYBFSTree}
\end{algorithm}

\begin{algorithm}[ht]
	\DontPrintSemicolon
	\SetKwInOut{Input}{Input}\SetKwInOut{Output}{Output}
	\Input   {Graph $G(V, E) $, activation set $A$ and list $L$.}
	\Output  {Highest Influenced Vertex $w$.}
	\SetKwFunction{FMain}{\HPT}
	\SetKwProg{Fn}{Function}{}
	\Fn{\FMain{$G,A,L$}}{
		\Begin{
            $Q \gets \phi $\; \tcp{Q is a Queue}
            $A \gets \phi $\;
            $Q.push(u), \; D_b[u]=u,\; VIS[u] = True, \; \forall u\in L $\; 
            \tcp{$VIS$ is an array that will keep track of visited vertices. }
            $W[u]=deg(u), \; VIS[u] = False, \; \forall u\in L $\;
            $max \gets -1$\;
            $w \gets -1$\;
            \While{$Q$ is not empty}{
                  $u \gets  Q.dequeue()$\;
                  
                  \For{$v \in N(u)$}{
                     \If{$VIS[v]=False$ and $v \notin A $}{
                        $D_b[v] \gets D_b[u]$\;
                        $W[v] \gets W[v] + W[u]$\;
                        $VIS[v] \gets True$\;
                        $Q.enqueue(v)$\;
                     }
                  }
                  \If{$max < W[v] $}{
                   $max \gets W[v]$\;
                   $w \gets D_b[u]$\;
                  }
            }
             \KwRet{$w$}
		}		
	}
	\caption{Get potential vertex using BFS-tree.}
	\label{algo:PotentialNodeBYBFSTree}
\end{algorithm}
As given in algorithm~\ref{algo:PotentialNodeBYBFSTree}, the inputs for the \HPT method are \G{}, $A$, and $L$. The method uses the queue data structure to find the heaviest BFS tree. The initial step is  to en-queue each vertex $u \in L$ to queue $Q$ and initialize $W[v]=0$ $ \forall v\in V$. 
Do de-queue $u$ from queue $Q$. For each unmarked vertex $v \in N(u) \backslash A$, update $W[v] $  by $ W[v] + W[u]$, enqueue $v$ to  queue $Q$, $D_b[v]=u$ ( where $D_b$ array keeps track of the root $u$ that discovers the vertex $v$. ) and mark $v$ as visited.  If $W[v]> max$,  then update $w$ by $D_b[v]$ and $max=W[v]$. If $Q$ becomes empty, stop; otherwise, repeat. In the last step, The \HPT method returns the vertex $w \in L$ associated with the heaviest BFS tree.
\subsection{BFS and  DFS Greedy Heuristics }
\label{subsec:algothree}
\citeauthor{2023tieredInf} proposed two heuristics, BFS-GREEDY and DFS-GREEDY. We also apply the \EXTR technique on the seed set returned by these heuristics to improve the seed set.


\section{Result and discussion}
\label{sec:resultanddiscussion}
We implemented our algorithms on the Ubuntu Operating System, and the hardware specifications are the processor Intel Core$ {TM}$ i7-8700CPU@3.2Ghz and 16GB RAM. 

\begin{figure}
    \centering
    \includegraphics[width=100mm,scale=0.3]{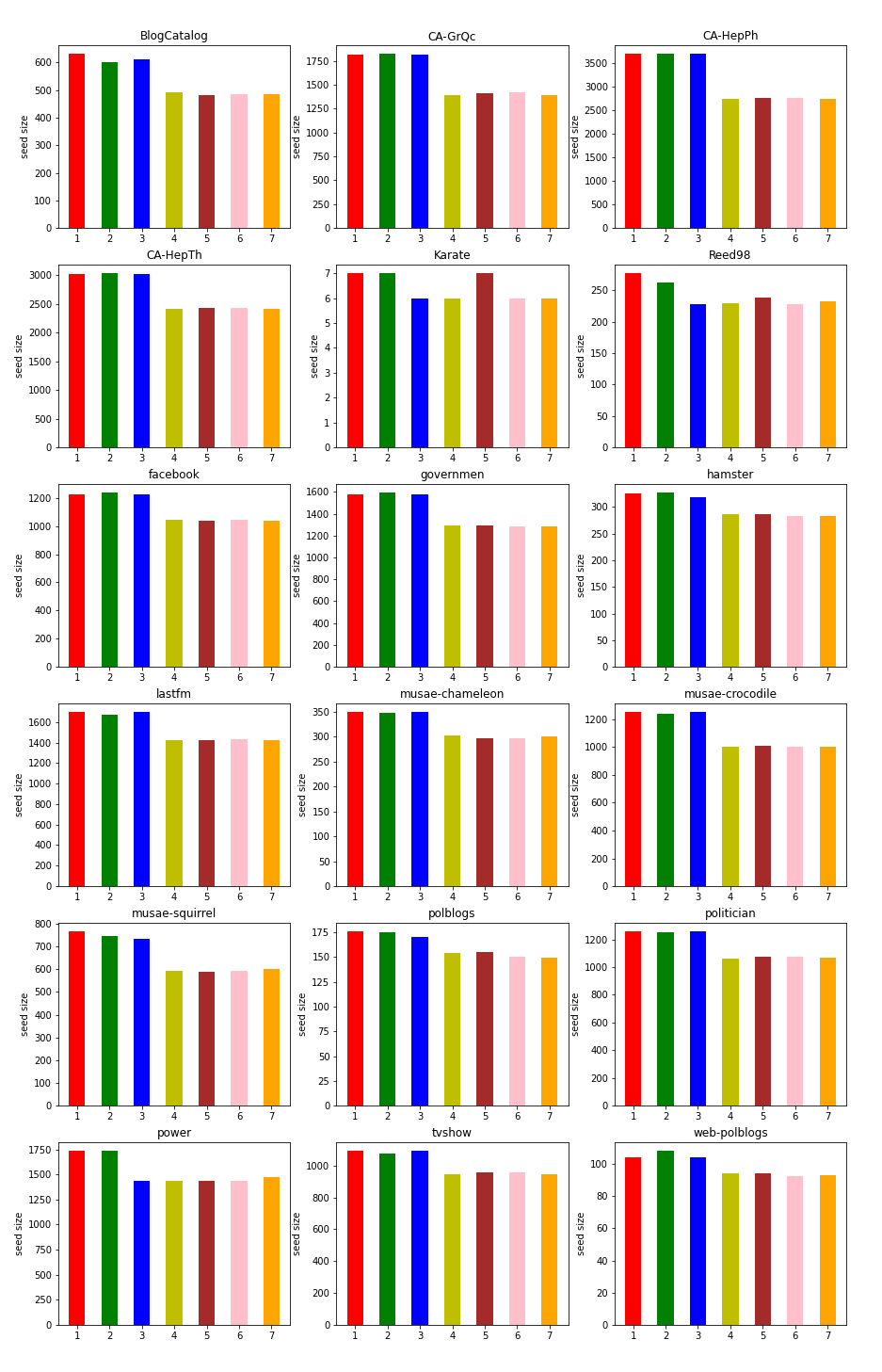}
    \caption{When propagation range $p=3$, comparison of the results between the proposed algorithms and ~\cite{2023tieredInf}. Algorithm numbers 1 (DFS-GREEDY~\cite{2023tieredInf}) and 2 (BFS-GREEDY~\cite{2023tieredInf}) are proposed by~\cite{2023tieredInf}, and we are proposing algorithm numbers 3 (\DFSGDEXT{}), 4 (\BFSGDEXT), 5(\BB), 6(\AM). The algorithm numbers are denoted on the X-axis of the bars,  and the Y-axis denotes the size of the seed set.}
    \label{fig:figureresultcomaresion}
\end{figure}

\begin{figure}
    \centering
    \includegraphics[width=100mm,scale=0.20]{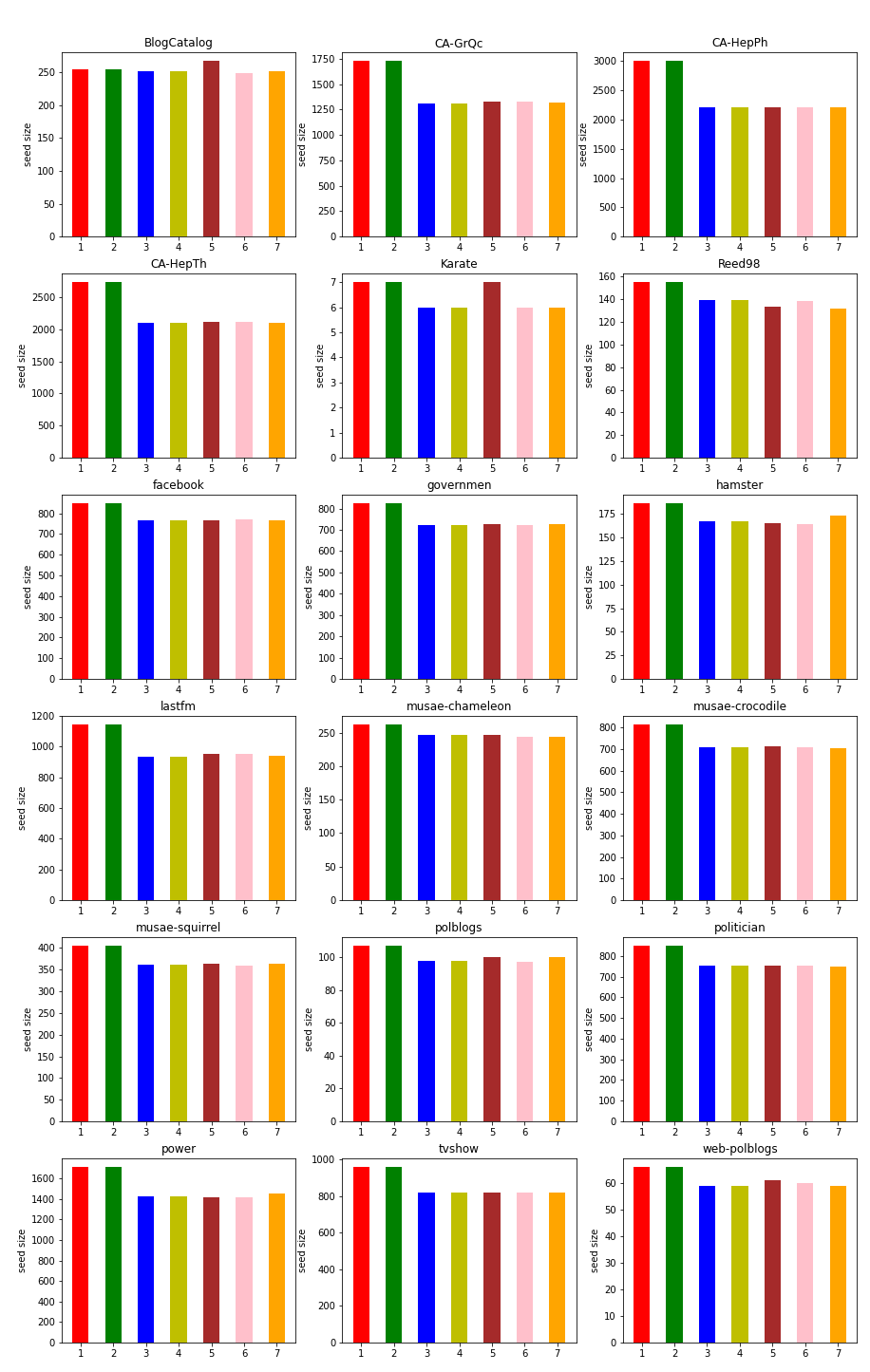}
    \caption{When propagation rang $p=diameter(G)$, the comparison of the results between the proposed algorithms and ~\cite{2023tieredInf}. Algorithm numbers 1 (DFS-GREEDY~\cite{2023tieredInf}) and 2 (BFS-GREEDY~\cite{2023tieredInf}) are proposed by~\cite{2023tieredInf}, and we are proposing algorithm numbers 3 (\DFSGDEXT{}), 4 (\BFSGDEXT), 5(\BB), 6(\AM). The algorithm numbers are denoted on the X-axis of the bars,  and the Y-axis denotes the size of the seed set.}
    \label{fig:evengilism}
\end{figure}

For comparison purposes, we set the parameters used in the algorithms as the propagation range $P=3$  and $P=diameter(G)$, the activation threshold $\alpha=0.6$, and the influence threshold $\theta = 0.4$.  The proposed algorithms are compared with recently published efficient algorithms by ~\citeauthor{2023tieredInf}\cite{2023tieredInf}. The sources of datasets are  network repository \cite{nr}, SNAP dataset\cite{snapnets}, social networks~\cite{reza2009social}, and data collected by \citeauthor{2015MarkNewMan}\cite{2015MarkNewMan}.  The results are shown for propagation range $p = 3$ in Figure~\ref{fig:figureresultcomaresion} and Table~\ref{table:whenp=3}. The results for  p=diameter are shown in Figure\ref{fig:evengilism} and Table~\ref{table:whenp=n}. In the tables, the  algorithms from \cite{2023tieredInf} are referred to as DFS-GREEDY (DFS-GD) and BFS-GREEDY (BFS-GD) while the proposed heuristics DFS-GREEDY-PRUNNING as DFS-PRUN, BFS-GREEDY-PRUNNING as BFS-PRUN, BACKBONE-BASED as BBH, AVERAGE-DEGREE as ADH and CLOSEST-FIRST as CFH.  

Average degree heuristic and backbone heuristics perform well for dense data sets like Karate~\cite{2015MarkNewMan}, Reed98~\cite{nr}, musae-squirrel~\cite{snapnets}, and Web-pol blogs as both the algorithms find spreader vertices based on the importance of the degree of inactive vertices.  Average Degree (ADH)  and Closest-First (CFH) heuristics perform well on dense graphs with a high average clustering coefficient and a high average degree. \BFSGDEXT and \BB heuristics work well on sparse graphs. The \EXTR works efficiently and effectively if the selected spreaders are in the neighborhood of each other. Our \BB algorithm uses Prim's algorithm to find the dominated path based on degree, which can activate more vertices. Therefore, all our proposed four heuristics are improving recent results given by~\citeauthor{2023tieredInf}\cite{2023tieredInf}.

\begin{table}[H]
    \centering
     \caption{Network properties of the data sets~\cite{2015MarkNewMan,snapnets,reza2009social,nr} used in this paper.}
     \scriptsize
    \begin{tabular}{p{2.5cm} p{1.5cm} p{1.5cm} p{1.5cm} p{2cm} p{1.5cm} p{1cm}}
    \hline
        Network & Nodes  & Edges & Density & Avg-triangles & Avg-Degree & Avg-CC \\
        \hline
        \hline
        power  &  4941  &  6594  &  0.0005  &  0.3953  &  2.669  &  0.08 \\ \hline
        BlogCatalog  &  10312  &  333983  &  0.0063  &  1631.6905  &  64.776  &  0.463 \\ \hline
        lastfm  &  7624  &  27806  &  0.001  &  15.9102  &  7.294  &  0.219 \\ \hline
        tvshow &  3892  &  17262  &  0.0023  &  67.13  &  8.871  &  0.374 \\ \hline
        web-polblogs  &  643  &  2280  &  0.011  &  14.0156  &  7.092  &  0.232 \\ \hline
        CA-HepTh  &  9877  &  25998  &  0.0005  &  8.6076  &  5.264  &  0.471 \\ \hline
        polblogs  &  1224  &  16718  &  0.0223  &  247.6544  &  27.317  &  0.32 \\ \hline
        Karate  &  34  &  78  &  0.139  &  3.9706  &  4.588  &  0.571 \\ \hline
        hamster  &  1858  &  12534  &  0.0073  &  27.0452  &  13.492  &  0.141 \\ \hline
        musae-crocodile  &  11631  &  170918  &  0.0025  &  160.8029  &  29.39  &  0.336 \\ \hline
        facebook  &  4039  &  88234  &  0.0108  &  1197.3335  &  43.691  &  0.606 \\ \hline
        CA-GrQc  &  5242  &  14496  &  0.0011  &  27.6192  &  5.531  &  0.53 \\ \hline
        Reed98  &  962  &  18812  &  0.0407  &  302.922  &  39.11  &  0.318 \\ \hline
        musae-squirrel &  5201  &  198493  &  0.0147  &  5534.8639  &  76.329  &  0.422 \\ \hline
        government &  7057  &  89455  &  0.0036  &  222.6955  &  25.352  &  0.411 \\ \hline
        musae-chameleon &  2277  &  31421  &  0.0121  &  451.9974  &  27.599  &  0.481 \\ \hline
        politician &  5908  &  41729  &  0.0024  &  88.6757  &  14.126  &  0.385 \\ \hline
        CA-HepPh  &  12008  &  118521  &  0.0016  &  839.0654  &  19.74  &  0.611 \\ \hline
    \end{tabular}
    \label{tab:graphs}
\end{table}

\begin{table}[H] 
\caption{Comparison of estimated seed set size when the propagation range is $p=3$.}
    \centering
    \begin{adjustbox}{angle=0} 
    \scriptsize
    \begin{tabular}{| m{2.6cm}|  >{\centering\arraybackslash}m{1.5cm} | >{\centering\arraybackslash} m{1.5cm} | >{\centering\arraybackslash}m{1.5cm} | >{\centering\arraybackslash}m{1.5cm} | >{\centering\arraybackslash}m{1.5cm} | >{\centering\arraybackslash} m{1.5cm}|  >{\centering\arraybackslash} m{1.5cm}|}
    \hline
     Network & DFS-GD~\cite{2023tieredInf} & BFS-GD~\cite{2023tieredInf} &DFS-PRUN & BFS-PRUN  & BBH & ADH& CFH \\ \hline 
     
     BlogCatalog &  631 &  600 &  613 &  493 &  \textbf{481} &  486 &  484\\
\hline
  CA-GrQc &  1814 &  1824 &  1814 &  1396 &  1417 &  1418 &  \textbf{1397}\\
\hline
  CA-HepPh &  3700 &  3702 &  3700 &  2755 &  2770 &  2766 &  \textbf{2746}\\
\hline
  CA-HepTh &  3014 &  3031 &  3014 &  2411 &  2428 &  2428 &  \textbf{2409}\\
\hline
  Karate &  7 &  7 &  6 &  6 &  6 &  \textbf{6} &  \textbf{6}\\
\hline
  Reed98 &  277 &  262 &  228 &  230 &  238 &  \textbf{228} &  232\\
\hline
  facebook &  1227 &  1237 &  1227 &  1046 &  \textbf{1037} &  1042 &  1038\\
\hline
  governmen &  1581 &  1594 &  1581 &  1295 &  1290 &  \textbf{1287} &  1288\\
\hline
  hamster &  326 &  327 &  318 &  286 &  286 &  284 &  \textbf{284}\\
\hline
  lastfm &  1697 &  1675 &  1697 &  \textbf{1423} &  1427 &  1430 &  1427\\
\hline
  musae-chameleon &  349 &  347 &  349 &  302 &  296 &  \textbf{296} &  300\\
\hline
  musae-crocodile &  1252 &  1242 &  1252 &  1005 &  1008 &  1005 &  \textbf{1003}\\
\hline
  musae-squirrel &  767 &  745 &  736 &  595 &  \textbf{591} &  595 &  602\\
\hline
  polblogs &  176 &  175 &  170 &  154 &  155 &  150 &  \textbf{149}\\
\hline
  politician &  1258 &  1251 &  1258 &  \textbf{1064} &  1075 &  1074 &  1071\\
\hline
  power &  1735 &  1735 &  1439 &  1439 &  1438 &  \textbf{1435} &  1475\\
\hline
  tvshow &  1092 &  1076 &  1092 &  \textbf{943} &  957 &  957 &  947\\
\hline
  web-polblogs &  104 &  108 &  104 &  94 &  94 &  \textbf{92} &  93\\
\hline

    \end{tabular}
    \label{table:whenp=3}
    \end{adjustbox}
\end{table}

\begin{table}[H] 
\caption{Comparison of estimated seed set size when the propagation range equals the graph's diameter.}
    \centering
    \begin{adjustbox}{angle=0} 
    \scriptsize
    \begin{tabular}{| m{2.6cm}|  >{\centering\arraybackslash}m{1.5cm} | >{\centering\arraybackslash} m{1.5cm} | >{\centering\arraybackslash}m{1.5cm} | >{\centering\arraybackslash}m{1.5cm} | >{\centering\arraybackslash}m{1.5cm} | >{\centering\arraybackslash} m{1.5cm}|  >{\centering\arraybackslash} m{1.5cm}|}
    \hline
     Network & DFS-GD~\cite{2023tieredInf} & BFS-GD~\cite{2023tieredInf} &DFS-PRUN & BFS-PRUN  & BBH & ADH& CFH \\ \hline 
      BlogCatalog &  254 &  254 &  252 &  252 &  267 &  \textbf{249} &  252\\
\hline
  CA-GrQc &  1727 &  1727 &  1312 &  \textbf{1312} &  1326 &  1326 &  1318\\
\hline
  CA-HepPh &  3004 &  3004 &  2205 &  \textbf{2205} &  2216 &  2216 &  2208\\
\hline
  CA-HepTh &  2731 &  2731 &  2095 &  \textbf{2095} &  2109 &  2109 &  2100\\
\hline
  Karate &  7 &  7 &  6 &  6 &  6 &  6 &  \textbf{6}\\
\hline
  Reed98 &  155 &  155 &  139 &  139 &  133 &  138 &  \textbf{132}\\
\hline
  facebook &  848 &  848 &  768 &  768 &  768 &  769 &  \textbf{767}\\
\hline
  governmen &  824 &  824 &  723 &  \textbf{723} &  729 &  724 &  725\\
\hline
  hamster &  186 &  186 &  167 &  167 &  165 &  \textbf{164} &  173\\
\hline
  lastfm &  1143 &  1143 &  934 &  \textbf{934} &  953 &  952 &  942\\
\hline
  musae-chameleon &  262 &  262 &  246 &  246 &  246 &  244 &  \textbf{244}\\
\hline
  musae-crocodile &  814 &  814 &  710 &  710 &  713 &  710 &  \textbf{704}\\
\hline
  musae-squirrel &  405 &  405 &  362 &  362 &  364 &  \textbf{360} &  363\\
\hline
  polblogs &  107 &  107 &  98 &  98 &  100 &  \textbf{97} &  100\\
\hline
  politician &  849 &  849 &  752 &  752 &  751 &  752 &  \textbf{747}\\
\hline
  power &  1711 &  1711 &  1421 &  1421 &  1419 &  \textbf{1419} &  1450\\
\hline
  tvshow &  960 &  960 &  819 &  \textbf{819} &  822 &  822 &  820\\
\hline
  web-polblogs &  66 &  66 &  59 &  59 &  61 &  60 &  \textbf{59}\\
\hline

    \end{tabular}
    \label{table:whenp=n}
    \end{adjustbox}
\end{table}


\section{Conclusion} \label{sec:concl}
Due to the importance of social media networks in daily life, this paper studies the influence maximization problem with propagation range. 
If a vertex receives the same information sufficient times from its neighbors, the vertex in the network becomes influenced. Similarly, an influenced vertex in the network starts spreading information if the vertex receives the same information from enough neighbors. Indeed, information originating from a source does not flow continuously. So, the influence model includes the propagation range of information from the originating vertex.  This paper proposes heuristics based on backbone-based heaviest paths and the average degree of non-activated vertices. The proposed heuristics and the pruning techniques give improved seed sets compared to existing heuristics. Applying genetic algorithms, particle swarm optimization, and other metaheuristic techniques to this problem is an interesting future direction. 
\bibliographystyle{plainnat}
\bibliography{myrefs}

\begin{thebibliography}{20}
\providecommand{\natexlab}[1]{#1}
\providecommand{\url}[1]{\texttt{#1}}
\expandafter\ifx\csname urlstyle\endcsname\relax
  \providecommand{\doi}[1]{doi: #1}\else
  \providecommand{\doi}{doi: \begingroup \urlstyle{rm}\Url}\fi

\bibitem[Alla and Kare(2023)]{alla2023opinion}
Leela~Srija Alla and Anjeneya~Swami Kare.
\newblock Opinion maximization in signed social networks using centrality
  measures and clustering techniques.
\newblock In \emph{Distributed Computing and Intelligent Technology: 19th
  International Conference, ICDCIT 2023, Bhubaneswar, India, January 18--22,
  2023, Proceedings}, pages 125--140. Springer, 2023.

\bibitem[Bhattacharya et~al.(2022)Bhattacharya, Das, and Dev]{k-center}
Binay Bhattacharya, Sandip Das, and Subhadeep~Ranjan Dev.
\newblock The weighted k-center problem in trees for fixed k.
\newblock \emph{Theoretical Computer Science}, 906:\penalty0 64--75, 2022.
\newblock ISSN 0304-3975.
\newblock \doi{https://doi.org/10.1016/j.tcs.2022.01.005}.
\newblock URL
  \url{https://www.sciencedirect.com/science/article/pii/S0304397522000135}.

\bibitem[Bonato et~al.(2016)Bonato, Janssen, and Roshanbin]{Bonato2016burn}
Anthony Bonato, Jeannette Janssen, and Elham Roshanbin.
\newblock How to burn a graph.
\newblock \emph{Internet Mathematics}, 12\penalty0 (1-2):\penalty0 85--100,
  2016.

\bibitem[Chen(2009)]{chen2009approximability}
Ning Chen.
\newblock On the approximability of influence in social networks.
\newblock \emph{SIAM Journal on Discrete Mathematics}, 23\penalty0
  (3):\penalty0 1400--1415, 2009.

\bibitem[Cordasco et~al.(2018)Cordasco, Gargano, Rescigno, and
  Vaccaro]{cordasco2018evangelism}
Gennaro Cordasco, Luisa Gargano, Adele~Anna Rescigno, and Ugo Vaccaro.
\newblock Evangelism in social networks: Algorithms and complexity.
\newblock \emph{Networks}, 71\penalty0 (4):\penalty0 346--357, 2018.

\bibitem[Cordasco et~al.(2019)Cordasco, Gargano, and
  Rescigno]{cordasco2019active}
Gennaro Cordasco, Luisa Gargano, and Adele~A Rescigno.
\newblock Active influence spreading in social networks.
\newblock \emph{Theoretical Computer Science}, 764:\penalty0 15--29, 2019.

\bibitem[Gautam et~al.(2022)Gautam, Kare, and Bhavani]{gautam2022faster}
Rahul~Kumar Gautam, Anjeneya~Swami Kare, and S.~Durga Bhavani.
\newblock Faster heuristics for graph burning.
\newblock \emph{Applied Intelligence}, pages 1--11, 2022.

\bibitem[Gautam et~al.(2023)Gautam, Kare, and Durga~Bhavani]{2023RahulCent}
Rahul~Kumar Gautam, Anjeneya~Swami Kare, and S.~Durga~Bhavani.
\newblock Centrality measures based heuristics for perfect awareness problem
  in social networks.
\newblock In Raghava Morusupalli, Teja~Santosh Dandibhotla, Vani~Vathsala
  Atluri, David Windridge, Pawan Lingras, and Venkateswara~Rao Komati, editors,
  \emph{Multi-disciplinary Trends in Artificial Intelligence}, pages 91--100,
  Cham, 2023. Springer Nature Switzerland.

\bibitem[Kazemzadeh et~al.(2022)Kazemzadeh, Safaei, and
  Mirzarezaee]{kazemzadeh2022influence}
Farzaneh Kazemzadeh, Ali~Asghar Safaei, and Mitra Mirzarezaee.
\newblock Influence maximization in social networks using effective community
  detection.
\newblock \emph{Physica A: Statistical Mechanics and its Applications},
  598:\penalty0 127314, 2022.

\bibitem[Kempe et~al.(2003)Kempe, Kleinberg, and Tardos]{kempe2003maximizing}
David Kempe, Jon Kleinberg, and {\'E}va Tardos.
\newblock Maximizing the spread of influence through a social network.
\newblock In \emph{Proceedings of the ninth ACM SIGKDD international conference
  on Knowledge discovery and data mining}, pages 137--146, 2003.

\bibitem[Leskovec and Krevl(2014)]{snapnets}
Jure Leskovec and Andrej Krevl.
\newblock {SNAP Datasets}: {Stanford} large network dataset collection.
\newblock \url{http://snap.stanford.edu/data}, June 2014.

\bibitem[Liang et~al.(2023)Liang, He, Du, and Xu]{2023targetedCompt}
Ziwei Liang, Qiang He, Hongwei Du, and Wen Xu.
\newblock Targeted influence maximization in competitive social networks.
\newblock \emph{Information Sciences}, 619:\penalty0 390--405, 2023.

\bibitem[Nazeri et~al.(2022)Nazeri, Mollahosseini, and
  Izadi]{geneticGraphBurning}
Mahdi Nazeri, Ali Mollahosseini, and Iman Izadi.
\newblock A centrality based genetic algorithm for the graph burning problem.
\newblock \emph{Available at SSRN 4224314}, 2022.

\bibitem[Newman(2015)]{2015MarkNewMan}
Mark Newman.
\newblock Network data.
\newblock \url{http://www-personal.umich.edu/~mejn/netdata/}, 2015.

\bibitem[Pereira et~al.(2021)Pereira, de~Rezende, and
  de~Souza]{pereira2021effective}
Felipe de~C Pereira, Pedro~J de~Rezende, and Cid~C de~Souza.
\newblock Effective heuristics for the perfect awareness problem.
\newblock \emph{Procedia Computer Science}, 195:\penalty0 489--498, 2021.

\bibitem[Qiang et~al.(2023)Qiang, Pasiliao, and Zheng]{2023tieredInf}
Zhecheng Qiang, Eduardo~L Pasiliao, and Qipeng~P Zheng.
\newblock Target set selection in social networks with tiered influence and
  activation thresholds.
\newblock \emph{Journal of combinatorial optimization}, 45:117, 2023.
\newblock \doi{https://doi.org/10.1007/s10878-023-01023-8}.

\bibitem[Reza and Huan(2009)]{reza2009social}
Zafarani Reza and Liu Huan.
\newblock Social computing data repository.
\newblock \url{http://datasets.syr.edu/pages/datasets.html}, 2009.

\bibitem[Rossi and Ahmed(2015)]{nr}
Ryan~A. Rossi and Nesreen~K. Ahmed.
\newblock The network data repository with interactive graph analytics and
  visualization.
\newblock In \emph{AAAI}, 2015.
\newblock \url{https://networkrepository.com}.

\bibitem[Zareie and Sakellariou(2022)]{rumorMinimization}
Ahmad Zareie and Rizos Sakellariou.
\newblock Rumour spread minimization in social networks: A source-ignorant
  approach.
\newblock \emph{Online Social Networks and Media}, 29:\penalty0 100206, 2022.
\newblock ISSN 2468-6964.
\newblock \doi{https://doi.org/10.1016/j.osnem.2022.100206}.
\newblock URL
  \url{https://www.sciencedirect.com/science/article/pii/S2468696422000106}.

\bibitem[Šimon et~al.(2019)Šimon, Huraj, Luptáková, and
  Pospíchal]{Simon2019heuristic}
Marek Šimon, Ladislav Huraj, Iveta Luptáková, and Jiří Pospíchal.
\newblock Heuristics for spreading alarm throughout a network.
\newblock \emph{Applied Sciences}, 9:\penalty0 3269, 2019.

\end{thebibliography}
\end{document}